\documentclass[usenatbib]{mn2e}
\usepackage[dvips]{graphicx}
\usepackage{epstopdf}
\def\CN2{\mbox{$C_N^2 \ $}}

\title[Mesoscale OT Simulations at Dome C]{Mesoscale Optical Turbulence simulations at Dome C}
\author[F. Lascaux et al.]{F. Lascaux,$^{1}$\thanks{E-mail:
     lascaux@arcetri.astro.it; masciadri@arcetri.astro.it} E. Masciadri$^1$\footnotemark[1],
S. Hagelin$^{1, 2}$ and J. Stoesz$^1$ \\ $^1$INAF Osservatorio Astrofisico
di Arcetri, Largo Enrico Fermi 5, I-501 25 Florence, Italy\\
$^2$Uppsala Universitet, Department of Earth Sciences, Villav\"agen 16,
S-752 36 Uppsala, Sweden}

\begin{document}
\label{firstpage}
\date{Accepted 2009 ??? ??, Received 2009 ??? ??; in original form
2009 ??? ??}  
\pagerange{\pageref{firstpage}--\pageref{lastpage}}
\pubyear{2009}

\maketitle

\begin{abstract}
These last years ground-based astronomy has been looking towards Antarctica, 
especially its summits and the internal continental plateau where the optical 
turbulence appears to be confined in a shallow layer close to the icy surface.
Preliminary measurements have so far indicated pretty good value for the seeing 
above 30-35~m: 0.36" \citep{ag}, 0.27" \citep{la} and 0.3" \citep{tr} at Dome C.
Site testing campaigns are however extremely expensive, instruments provide only 
local measurements and atmospheric modeling might represent a step ahead towards 
the search and selection of astronomical sites thanks to the possibility to 
reconstruct 3D $C_N^2$ maps over a surface of several kilometers.
The Antarctic Plateau represents therefore an important benchmark test to evaluate
the possibility to discriminate sites on the same plateau.
Our group \citep{ha} has proven that the analyses from the ECMWF global model do 
not describe with the required accuracy the antarctic boundary and surface layers 
in the plateau.
A better description could be obtained with a mesoscale meteorological model.
The mesoscale model Meso-Nh has proven to be reliable in reproducing 3D
maps of optical turbulence above mid-latitude astronomical 
sites \citep{m1,m2,m5,m3}. In this paper we study the ability of the Meso-Nh model in reconstructing the 
meteorological parameters as well as the optical turbulence above Dome C with different 
model configurations (monomodel and grid-nesting). 
We concentrate our attention on the
model abilities in reproducing the optical turbulence surface layer thickness ($h_{sl}$) and the 
integral of the $\CN2$ in the free atmosphere and in the surface layer. It is worth to highlight 
that these are the first estimates ever done so far
with a mesoscale model of the optical turbulence above the internal Antarctic Plateau.

\end{abstract}


\section{Introduction}
The internal Antarctic Plateau is, at present, a site of potential great interest 
for astronomical applications. 
The extreme low temperatures, the dryness, the typical high altitude of the 
internal Antarctic Plateau (more than 2500~m), joint to the fact that the optical 
turbulence seems to be concentrated in a thin surface layer whose thickness is of 
the order of a few tens of meters do of this site a place in which, potentially,
we could achieve astronomical observations otherwise possible only by space.
In spite of the exciting first results \citep{la,ar,tr} the 
effective gain that astronomers might achieve from ground-based astronomical 
observation from this location still suffers from serious uncertainties and doubts 
that have been pointed out in previous work \citep{m4,gm,ha,stoesz08}.
A better estimate of the properties of the optical turbulence above the internal 
Antarctic Plateau can be achieved with both dedicated measurements done 
simultaneously with different instruments and simulations provided by 
atmospheric models.
Simulations offer the advantage to provide volumetric maps of the optical 
turbulence ($C_N^2$) extended on the whole internal plateau and, ideally, to 
retrieve comparative estimates in a relative short time and homogeneous way on 
different places of the plateau.
In a previous paper \citep{ha} our group performed a detailed analysis of the 
meteorological parameters from which the optical turbulence depends on, provided by
the General Circulation Model (GCM) of the European Center for Medium-range Weather
Forecasts (ECMWF).
In that work we quantified the accuracy of the ECMWF estimates of all the major 
meteorological parameters and, at the same time, we pointed out which are the 
limitations of the GCMs.
In contexts in which the GCMs fail, mesoscale models can supply more accurate information 
because they are conceived to reconstruct phenomena 
that develop at a too small spatial and temporal scale to be described 
by a GCM.
In spite of the fact that mesoscale models can attain higher resolution than the 
GCMs, some parameters, such as the optical turbulence, are not explicitly resolved 
but are parameterized, i.e. the fluctuations of the microscopic physical quantities
are expressed as a function of the corresponding macroscopic quantities averaged on
a larger spatial scale (cell of the model).
For classical meteorological parameters the use of a mesoscale model should be 
useless if GCMs such as the one of the ECMWF could provide estimate with equivalent
level of accuracy.
For this reason the Hagelin et al. paper (2008) has been a first step towards the 
exploitation of the mesoscale Meso-Nh model above the internal Antarctic Plateau.
In that study we retrieved an exhaustive characterization of all the meteorological parameters 
from the ECMWF analyses (wind field, 
potential temperature, absolute temperature...) and, at the same time, we defined the ECMWF's analyses limitations:
we concluded that in the first 10-20~m, the ECMWF analyses show a discrepancy with 
respect to measurements of the order of 2-3~m.s$^{-1}$ for the wind speed and of 
4-5 K for the temperature.
\begin{figure*}
 \includegraphics[width=0.8\textwidth]{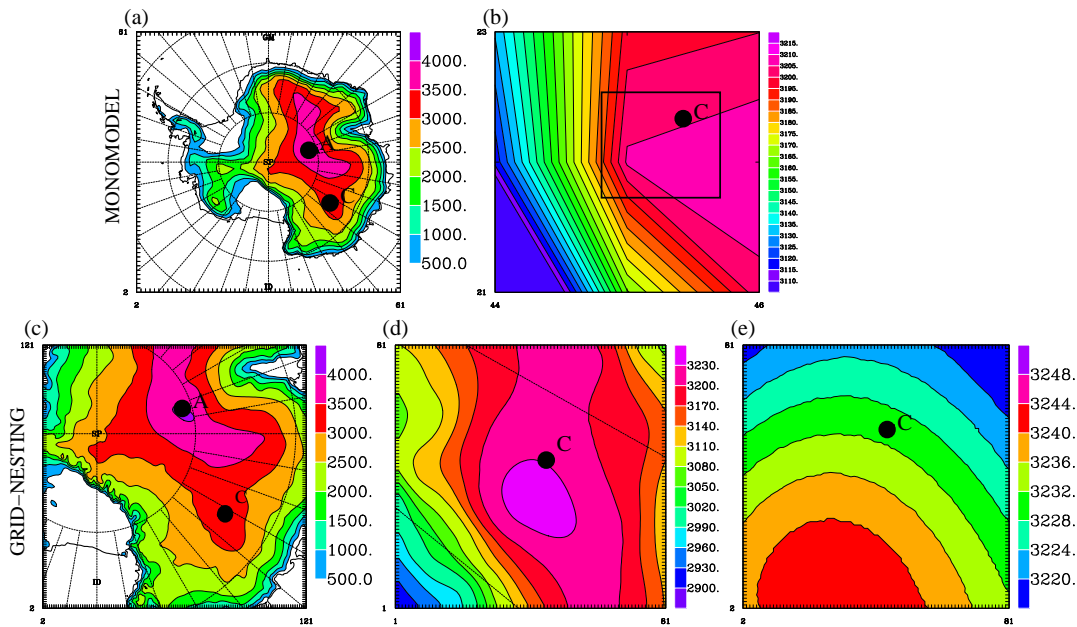}
 \caption{Orography of Antarctica as seen by the Meson-Nh model (polar
 stereographic projection) in
 (a) the monomodel simulation, with horizontal resolution ${\Delta}X$=100~km;
 (b) zoom of (a) above the Dome C area (the black square represents the same area
 as in (e));
 (c) the largest domain of the grid-nested simulation with ${\Delta}X$=25~km;
 (d) the second domain of the grid-nested simulation with ${\Delta}X$=5~km;
 (e) the innermost domain of the grid-nested simulation with  ${\Delta}X$=1~km.
 The dot labeled 'C' is located at the Concordia Station. The dot labeled 'A' is
 the Dome A. Altitude in meter (m).}
{\label{fig:oro}}
\end{figure*}
\par
The Meso-Nh model has been proven to be reliable in reproducing 3D
maps of optical turbulence \citep{m1,m2,m8} and it has been statistically validated \citep{m5,m3,m4} above mid-latitude astronomical 
sites\footnote{See details in the next section}. Preliminary tests concerning the optimization of the model configuration and sensitivity to the horizontal and the vertical resolution have already been conducted by our 
team \citep{lf} for the internal Antarctic plateau. 
In this paper we intend to quantify the performances of the model above this peculiar environment. More precisely, our goals are:
\begin{itemize}
\item[-] to compare the performances of the mesoscale Meso-Nh model and the ECMWF 
GCM in reconstructing wind speed and absolute temperature (main meteorological 
parameters from which the optical turbulence depends on) with respect to the
 measurements. This analysis will quantify the performances of the Meso-Nh model with respect 
to the GCM from the ECMWF.
\item[-] to perform simulations of the optical turbulence above Dome C (75$^{\circ}$06'04"S, 123$^{\circ}$20'48"E) 
employing different model configurations and compare the typical simulated thickness of the 
surface layers well as the seeing in the free atmosphere with the one measured by \cite{tr} (hereafter TR2008).
In this way we aim to establish which configuration is necessary to reconstruct 
correctly the $C_N^2$. In summary we aim to validate the Meso-Nh model on the Antarctic site.
\end{itemize}
The two issues: (1) the surface layer thickness $h_{sl}$ and  (2) the typical seeing in the free atmosphere are certainly the two main 
features that might get this place on the Earth extremely appealing for astronomers and it might be extremely useful to have an independent confirmation from models of the typical values measured on the site.
This study is focused on the winter season. In Section 2 we present the Meso-Nh model and the different configurations that 
were used to perform numerical weather simulations above the internal antarctic 
plateau.
Section 3 is devoted to a statistical comparison of standard meteorological parameters
(wind speed and temperature) deduced from Meso-Nh simulations, ECMWF analyses 
and radiosoundings.
In Section 4 we present the results of the computation with Meso-Nh of the surface 
layer thickness for 15 nights in winter time and a comparison with the observed 
surface layer thickness from TR2008.
Finally conclusions are drawn in Section 5.

\section{Model numerical set-up}
Meso-Nh \citep{laf} is the non-hydrostatic mesoscale research model 
developed jointly by M\'et\'eo-France and Laboratoire d'A\'erologie.
\par
It can simulate the temporal evolution of the three-dimensional atmospheric flow 
over any part the globe.
The prognostic variables forecasted by this model are the three cartesian 
components of the wind $u$, $v$, $w$, the dry potential temperature $\Theta$, the 
pressure $P$, the turbulent kinetic energy $TKE$.
\par
The system of equation is based upon an anelastic formulation 
allowing for an effective filtering of acoustic waves.
A Gal-Chen and Sommerville\citet{gcs} coordinate on the vertical and a C-grid 
in the formulation of Arakawa and Messinger\citet{am} for the spatial 
digitalization is used.
The temporal scheme is an explicit three-time-level leap-frog scheme with a time 
filter \citep{as}.
The turbulent scheme is a one-dimensional 1.5 closure scheme \citep{cux} with the 
Bougeault and Lacarr\`ere\citet{bl} mixing length.
The surface exchanges are computed in an externalized surface scheme (SURFEX) 
including different physical packages, among which ISBA \citep{np} for vegetation.
\par
Masciadri et al. (1999a,b) implemented the optical turbulence package
to be able to forecast also the optical turbulence ($C_N^2$ 3D maps) and all the 
astroclimatic parameters deduced from the $C_N^2$. 
We will refer to the 'Astro-Meso-Nh code' to indicate this package.
To compare simulations with measurements the integrated astroclimatic parameters 
are calculated integrating the  $C_N^2$ with respect to the zenith in the 
Astro-Meso-Nh code. 
The parameterization of the optical turbulence and the reliability of the 
Astro-Meso-Nh model have been proved in successive studies in which simulations 
have been compared to measurements provided by different instruments \citep{m8,m3,m4}.
This has been achieved thank to a dedicated calibration procedure that has been proposed and validated by the same authors \citep{m5}.
\par
The atmospheric Meso-Nh model is conceived for research development and for this 
reason is in constant evolution. 
One of the major advantages of Meso-Nh that was not available at the time of the 
Masciadri's studies is that it allows now for the use of  the interactive 
grid-nesting technique \citep{st}.
This technique consists in using different imbricated domains with increasing 
horizontal resolutions with mesh-sizes that can reach 10 meters.
\par
We use in this study the Astro-Meso-Nh package,  implemented in the most recent version of the 
atmospheric Meso-Nh model. 
To facilitate the put in the context of this work, the differences that have been implemented in the model configuration 
with respect to the previous Masciadri's studies are listed here:

\begin{itemize}
\item A higher vertical resolution near the ground has been selected.
We still work with a logarithmic stretching near the ground up to 3.5~km but we 
start with a first grid point of 2~m (instead of 50~m) with 12 points in the first 
hundred meters.
This configuration has been allowed thanks to the 
extremely smooth orography of this region of the Earth. It is 
obviously preferable because it permits to better quantify the turbulence 
contribution that typically develops in the thin vertical slabs in the first hundred of meters above the 
internal Antarctic Plateau. 
Above 3.5~km the vertical resolution is constant and equal to ${\Delta}H$=600~m as 
well as in Masciadri's previous work.
The maximum altitude is 22 kilometers.
\item The grid-nesting (see Table~\ref{tab1}) is implemented with 3 imbricated domains allowing a maximum 
horizontal resolution of 1~km in a region around the Concordia Station (80~km $\times$ 80~km). 
\item The simulations are forced at synoptic times (every 6 hours) by analyses from the ECMWF.
This permits to perform a real forecast of the optical turbulence. 
To avoid misunderstandings, we highlight indeed that, as it has been extensively explained in previous studies (Masciadri et al. 2004, Masciadri \& Egner 2006), the Meso-Nh model 
has been used so far for simulations of the optical turbulence in a 
configuration permitting a quantification of the mean optical 
turbulence during a night and not a forecast of the optical turbulence. We perform therefore
a step ahead with respect to results obtained so far with the Astro-Meso-Nh code.
\end{itemize}
\par
In spite of the fact that the orographic morphology is almost flat above 
Antarctica, it is known that even a weak slope can be an important factor to induce a 
change in the wind speed at the surface in these regions.
The physics of the optical turbulence strongly depend on a delicate balance between
the wind speed and temperature gradients. 
In order to study the sensitivity of the model to the horizontal resolution and to identify 
which configuration provides more reliable estimates we
performed two sets of simulations with different model configurations.
\par
In the first configuration (that we will call {\it monomodel}) we used an 
horizontal resolution ${\Delta}X$=100~km covering the whole Antarctic continent 
(Figure \ref{fig:oro}a,b and Table \ref{tab1}).
We selected this configuration because it permits us to discuss, where it is possible, our results with respect to those obtained by \cite{seg} (hereafter SG2006) with the regional 
atmospheric model MAR above the Antarctic Plateau. In that case, indeed, the authors used this extremely low horizontal resolution that has the advantage to be cheap from a computational point of view. This model configuration permits fast simulations but  it is certainly necessary to verify that it is high enough to correctly resolve the most important features of the optical turbulence near the ground and in the high part of the atmosphere. 
\begin{table}
\caption{Meso-Nh model configuration. In the second column the  horizontal resolution $\Delta$X, in the third column the number of grid points and in the fourth column the horizontal surface covered by the model domain.}
\begin{tabular}{cccc}
\hline
 & $\Delta$X (km) & Grid Points & Surface (km)\\
\hline
Monomodel  & 100 & 60$\times$60 & 6000$\times$6000  \\
\hline
                    & 25& 120$\times$120 & 3000$\times$3000  \\
Grid-Nesting & 5 & 80$\times$80 &   400$\times$400\\
                     & 1 & 80$\times$80 & 80$\times$80  \\
\hline
\end{tabular}

\label{tab1}
\end{table}

In the second configuration (Table \ref{tab1}) we used the grid-nesting technique, more expensive from a computational point of view
but potentially more accurate in the reconstruction of the spatial distribution of the optical turbulence.
The grid-nested simulations involved three domains.
The largest domain covers all the Antarctic Plateau with 
120x120 points and it has a 25~km mesh-size (Figure \ref{fig:oro}c).
The second domain has a 5~km mesh-size, 80x80 points and is centered 
above the Dome C (Figure \ref{fig:oro}d).
The innermost domain has a 1~km mesh-size, 80x80 points and is centered above the 
Concordia Station area near the Dome C (Figure \ref{fig:oro}e). Due to the fact that the typical topography of the internal Antarctic Plateau is much smoother than what is typically observed at mid-latitude sites, it has been decided to use a maximum resolution of 1 km instead of 500 m as it has been done in all the previously Masciadri's cited studies.
\par
The use of high-resolution has one first major impact: as it can be seen in Fig.~\ref{fig:oro}, the Dome C area is more 
fairly reproduced in the grid-nested simulation than in the low horizontal 
resolution simulation (Figure \ref{fig:oro}b,e).
The altitude above mean sea level of the Concordia Station with high resolution 
is around 3230~m, whereas it is around 3200~m with the low resolution grid.

\section{Meso-Nh simulations: absolute temperature and wind speed}
\begin{figure*}
\begin{center}
\includegraphics[width=0.9\textwidth]{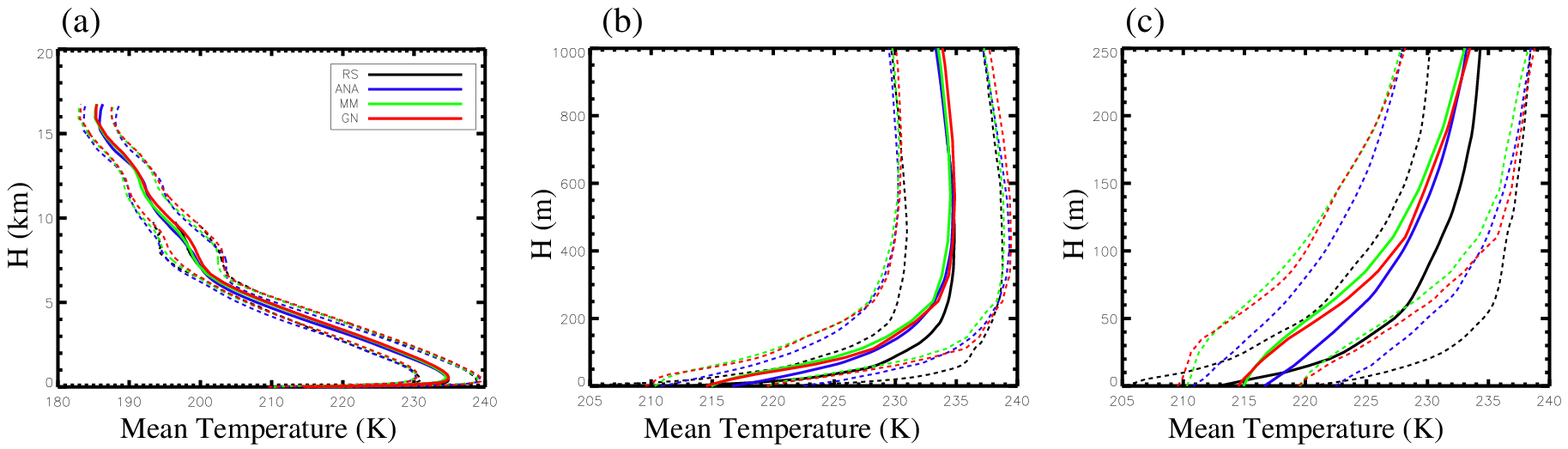}
\end{center}
\caption{Mean temperature profiles at the Concordia Station in the
Dome C area over 47 winter nights (bold lines) and the corresponding standard
deviation (dashed lines). Radiosoundings: black lines; ECMWF analyses: blue lines; 
Meso-Nh with low horizontal resolution (monomodel): green lines; Meso-Nh with 
high resolution (grid-nesting): red lines.
Profiles are displayed up to (a) 20~km, (b) 1~km and (c) 250~m above ground level. Units in Kelvin (K).}
{\label{fig:mean_t}}
\begin{center}
\includegraphics[width=0.9\textwidth]{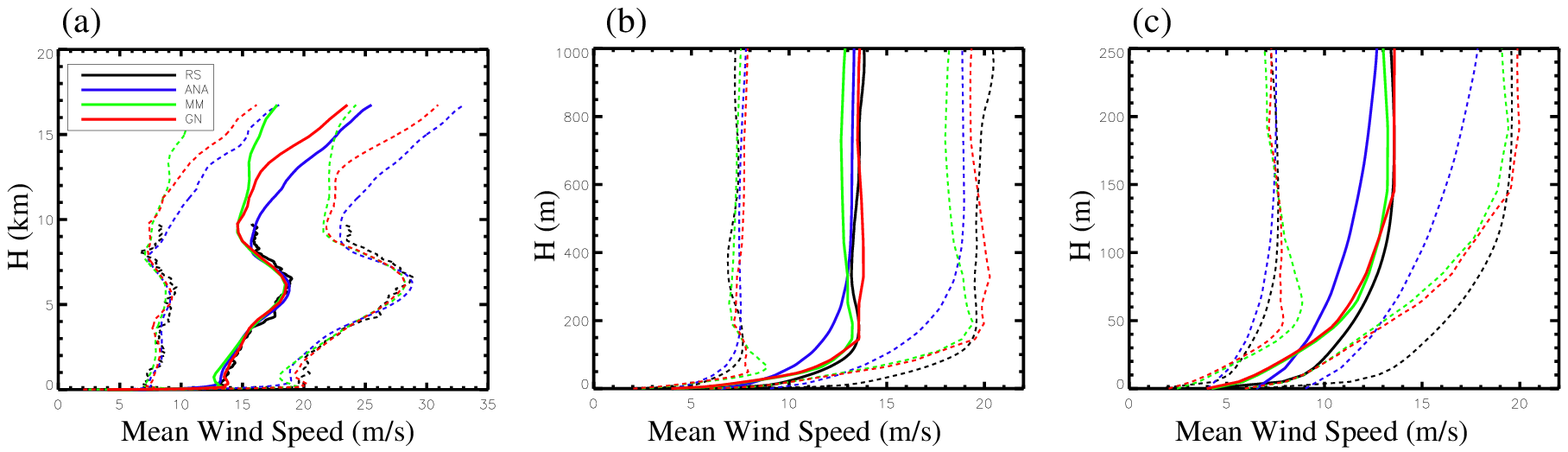}
\end{center}
\caption{Mean wind speed profiles at the Concordia Station in the Dome C area 
over 47 winter nights (bold lines) and the corresponding standard deviation (dashed
lines).
Radiosoundings: black lines; ECMWF analyses: blue lines; Meso-Nh with low horizontal resolution (monomodel): green lines; Meso-Nh with high-resolution (grid-nesting): red lines. 
Profiles are displayed up to (a) 20~km, (b) 1~km and (c) 250~m above ground level. Units in m.s$^{-1}$.}
{\label{fig:mean_w}}
\end{figure*}
The purpose of this section is to verify the performances of the mesoscale Meso-Nh 
model above the internal Antarctic Plateau and to verify if such a mesoscale model 
can provide a better estimate of the atmospheric flow than the GCM from the ECMWF.
\par
An important number of winter nights (47) were simulated with the Meso-Nh mesoscale
model.
We analyze here the key meteorologic parameters from which the optical turbulence 
depends on: the temperature and the wind speed.
Both configurations (low horizontal resolution monomodel, and high horizontal 
resolution grid-nesting) are tested and evaluated. The model is initialized with ECMWF analyses
extracted in the nearest grid point with respect to Dome C\footnote{Horizontal resolution of the ECMWF analyses: 
$\sim$0.5$^{\circ}$}.
All the simulations started at 00 UTC and were integrated for 12 hours.
Simulations outputs at 12 UTC are compared with measurements we retrieved from the site
({\it http://www.climantartide.it}) as well as with the analyses from the GCM of 
the ECMWF.
Every 6 hours we forced the simulations with the ECMWF analyses in order to avoid 
that the model diverges and/or correct the atmospheric flow as a function of the 
predictions at larger spatial scales. 
\par
In this section a statistical study of the wind and temperature profiles at 
Concordia Station, Dome C, is performed.
The 47 nights have been selected in June, July and August 2005 and July 2006.
For all the 47 nights selected, we respected the following criterion:
\begin{itemize}
\item A radiosounding is available  at the end of the simulation (at 12 UTC of 
the selected night) to perform comparisons between Meson-Nh outputs, 
ECMWF analyses and observations.
\item We selected nights in which, the corresponding radiosoundings launched at time t=t$_{0}$+12 h 
(with t$_{0}$ the initial time of the simulation) 
cover the longest path along the z-axis (perpendicular to the ground) before to explode.
It was impossible to collect in winter time 47 nights in which all the balloons 
reached 20~km.
The mean altitude reached by the selected balloons was about 10~km above ground 
level.
\end{itemize}

\subsection{Model validation: vertical profiles of temperature 
and wind speed}
Figures \ref{fig:mean_t} and \ref{fig:mean_w} show the mean vertical profiles 
of temperature and wind speed, respectively, computed for 47 winter nights from 
the two model configurations (low and horizontal resolution), the ECMWF analyses 
and the radiosoundings. 
The location of the profiles is Concordia Station, in the Dome C area.
All profiles have been interpolated on a regular 5~m vertical grid, in order to 
ease the comparison.
\begin{table}
 \centering
 \begin{minipage}{80mm}
 \caption{Mean values on 47 winter days at Concordia Station near Dome C of wind
speed and temperature at the surface level, for radiosoundings and ECMWF analyses.
Into brackets, the corresponding statistical error ($\sigma$/$\sqrt{N}$).}
 \begin{tabular}{|l|c|c|} 
 \hline
 \multicolumn{1}{l|}{} &  Radiosondes & ECMWF \\
 \hline
 Wind speed (m.s$^{-1}$) & 4.02 ($\pm$ 0.37) & 6.51 ($\pm$ 0.37) \\
 Temperature (K) & 212.90 ($\pm$ 1.11) & 216.64 ($\pm$ 0.85) \\
 \hline
 \label{tab:surf1}
 \end{tabular}
 \caption{Mean values on 47 days at Concordia Station near Dome C of wind speed and the 
temperature at the surface level, for the Meso-Nh simulations: 
1-MOD is for the low horizontal simulation with ${\Delta}X$=100~km and Grid-N is 
for the high horizontal grid-nested simulation with ${\Delta}X$=1~km for the 
innermost model centered above the Dome C area. 
Into brackets, the corresponding statistical error ($\sigma$/$\sqrt{N}$).}
 \begin{tabular}{|l|c||c|}
 \hline
 \multicolumn{1}{l|}{} & 1-MOD & Grid-N \\
 \hline
 Wind speed (m.s$^{-1}$) & 4.23 ($\pm$0.26) & 3.98 ($\pm$ 0.28) \\
 Temperature (K) & 214.92 ($\pm$ 0.68) & 214.50 ($\pm$ 0.72) \\      
 \hline
  \label{tab:surf2}
 \end{tabular} 
 \end{minipage} 
\end{table}

\subsubsection{Temperature}

The mean temperature profiles are very similar over the entire free atmosphere 
(Figure \ref{fig:mean_t}a). In the first kilometer the temperature gradients reconstructed 
by the Meso-Nh simulations (with high and low resolution) and the ECMWF 
analyses are not as pronounced as the one obtained with the radiosoundings. 
This means that the mesoscale model as well as the
General Circulation Model (ECMWF) reconstruct a slightly less stable atmosphere in this region 
even if the mesoscale model better approaches the observations trend. 
However, as it will be shown in the next section, the mesoscale model (Meso-Nh) provides a 
more accurate estimate of the surface temperature than what the ECMWF model can do.

\subsubsection{Wind speed}
Figure \ref{fig:mean_w} shows the mean wind speed during the 47 days, with the 
corresponding standard deviation.
From the ground up to 10~km, analyses and radiosoundings mean wind speed are well 
correlated.
Above 10~km the wind speed reconstructed by the ECMWF analyses is slightly larger than the 
one reproduced by Meso-Nh (monomodel and grid-nesting - left of figure 
\ref{fig:mean_w}).
It is hard to say whether the ECMWF analysis or the Meso-Nh simulation is the best 
since no mean value from the observations is available at this altitude.
Between 1~km and 10~km there are no major differences between the mesoscale model and the ECMWF.
Below 1~km it is well visible that Meso-Nh better reconstructs the strong wind shear than the ECMWF 
analyses up to achieve a more correlated wind speed value near the surface. 
At 150~m this difference is maximized: the wind speed provided by the ECMWF analyses is a bit too weak (12~m.s$^{-1}$ 
instead of 14~m.s$^{-1}$ in the observations). 
At the same altitude the Meso-Nh simulations give better results, with a mean wind 
profile perfectly correlated to the one measured by the radiosoundings. 
The improvement is even better in the case of the high-resolution model.
The difference between low horizontal and high horizontal simulations is more 
important above 12~km, with an increase in intensity of the wind more important in 
the high-resolution simulation.
These results match perfectly with 
a dedicated analysis that our group did \citep{ha} on a comparison between the 
wind speed provided by ECMWF analyses and radiosoundings near the surface.

\subsection{Model validation: the surface}
\label{meteo}

The mean values of the surface wind speed and absolute temperature at Dome C were computed for the 47 nights.
The results are reported in the Tables \ref{tab:surf1} and \ref{tab:surf2} (mean 
values for ECMWF analyses, radiosoundings, and Meso-Nh low and high horizontal 
resolution simulations, respectively). In another paper \citep{ha} our group showed that the radiosoundings in the first grid-point
perfectly match with measurements provided by the automatic weather stations (AWS).

\subsubsection{Temperature}

One can see that the ECMWF analyses, as already reported in a previous paper of 
our team \citep{ha} on a different data-set sample, are too warm at the surface (with a difference of almost 4 K,
table \ref{tab:surf1}) in winter with respect to the observations.
\par
The mean surface temperature simulated by Meso-Nh after 12 hours is 
closer to the observations than the ECMWF analyses.
The difference between ECMWF analyses and observed mean temperature is $\Delta$T$_{ecmwf,obs}$$=$3.74 K.
The mean surface temperature in the low-resolution simulation is $\Delta$T$_{mnh-low,obs}$$=$2.02 K higher 
than in the observations.
It is only $\Delta$T$_{mnh-high,obs}$$=$1.60 K higher for the grid nested simulation. This means
that the mesoscale model reconstructs a surface temperature that is typically a factor $\sim$2-2.5 (that is $\Delta$T$_{ecmwf,obs}$/$\Delta$T$_{mnh-high,obs}$) more accurate than the ECMWF analyses.
More over it can be seen that the surface temperature is better retrieved with the 
use of the high horizontal resolution (${\Delta}X$=1~km in the innermost domain) 
than the low horizontal resolution (${\Delta}X$=100~km).

\subsubsection{Wind speed}
Tables \ref{tab:surf1} and \ref{tab:surf2} show the mean values of the wind speed 
at the first interpolated point of the profiles.
The mean wind speed in the ECMWF analyses (6.51~m.s$^{-1}$) is higher than 
the observed wind speed (4.02~m.s$^{-1}$), thus a difference of ($\Delta$V$_{ecmwf,obs}$$=$
2.49~m.s$^{-1}$)\footnote{Difference from the \cite{ha} paper values are just due 
to the fact that in that paper, all the nights of the three months (June, July and 
August) were used while in this paper we simulated just 47 nights selected in two 
different years (2005 and 2006).
The statistical sample is therefore not the same.}.
Both mesoscale low and high horizontal simulations reproduce more accurately the surface wind speed than the
ECMWF analyses.
With a mesh-size of 100~km in Meso-Nh, the difference between the simulated and measured mean wind speed 
is of $\Delta$V$_{mnh-low,obs}$$=$ 0.21~m.s$^{-1}$.
The grid nested simulations (${\Delta}X$=1~km in the innermost domain) give even better results with a difference of $\Delta$V$_{mnh-high,obs}$$=$0.04 m.s$^{-1}$ only. This means
that the mesoscale model reconstructs a surface wind speed that is typically a factor $\sim$ 60 (that is $\Delta$V$_{ecmwf,obs}$/$\Delta$V$_{mnh-high,obs}$) more accurate than the ECMWF analyses.

The median value obtained with the mesoscale model MAR (SG2006) differs from the measurements, done on the same statistical sample, $\Delta$V$_{MAR,obs}$$=$ 0.9~m.s$^{-1}$. 
We conclude therefore that our simulations performed with Meso-Nh are a factor $\sim$ 4 
($\Delta$V$_{MAR,obs}$/$\Delta$V$_{mnh-low,obs}$) more accurate than those performed with MAR by SG2006 if we 
use the same horizontal resolution. 
They are a factor $\sim$ 22 ($\Delta$V$_{MAR,obs}$/$\Delta$V$_{mnh-high,obs}$) more accurate if we use the high 
horizontal resolution with Meso-Nh. 
We also note that in that paper is reported just the standard deviation ($\sigma$) and not the statistical error 
($\sigma$/$\sqrt{N}$) where N is the number of independent estimates. 
The former is appropriate to describe the dispersion of the data-set and the latter is useful to describe the 
precision of the estimate of the mean values and it is therefore more appropriate to quantify the performances 
of the simulations with respect to observations. 
This means that each standard deviation in Table 1 (SG2006) should be multiplied by 1/$\sqrt{N}$, 
where N=90 is the number of nights in three months. \newline

\section{Meso-Nh simulations: optical turbulence }
\begin{figure}
\begin{center}
\includegraphics[width=0.4\textwidth]{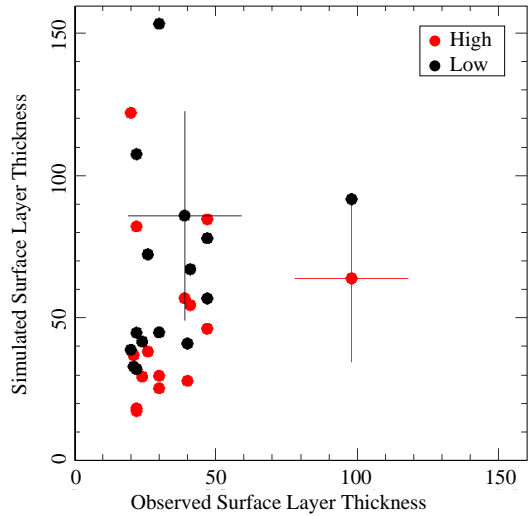}
\end{center}
\caption{Correlation plot between measured and simulated surface layer thicknesses (black: monomodel
configuration; red: grid-nested configuration). For the simulated values only the mean values between 12 UTC
and 16 UTC are considered.
For each configuration of the simulation (high and low horizontal resolution) the error bars are
reported for one point only (and are equal to $\sigma$). Units are in meter (m).}
{\label{fig:corr_slt}}
\end{figure}
We have seen in the previous section that the Meso-Nh model allows for a better 
forecasting of meteorological parameters such as wind speed and temperature than 
the analyses from the GCM of the ECMWF.
These thermodynamic parameters are very important for the forecasting of optical turbulence because 
the computation of the astroclimatological parameters (seeing $\varepsilon$, isoplanatic 
angle $\theta_{0}$, wavefront coherence time $\tau_{0}$) depend directly from them. 
As it has been highlighted in the introduction, the model has been run with the
Astro-Meso-Nh package that allows the prediction of 3D $\CN2$ maps.
The most important features that characterize the optical turbulence above the internal Antarctic Plateau are:
\begin{itemize}
\item the typical surface layer thickness
\item the median seeing in the free atmosphere
\item the median seeing in the whole atmosphere.
\end{itemize}
These three elements permit us to do a complete analysis of the optical turbulence developed above Dome C. Our tests are made on all the 15 nights in winter time for which measurements of the optical turbulence surface 
thickness, and partial seeing in different vertical slabs (free atmosphere seeing and total seeing) are available 
(see TR2008). 
Due to the fact that in the Trinquet et al. (2008) paper is available just the median $\CN2$ profile for the four seasons and 
not the $\CN2$ profile for each individual nights, we selected our sample taking all and only the nights belonging to the 
winter as defined by TR2008 i.e. between June 21 to September 21. This is the most interesting period for stellar astronomical applications.

\subsection{Surface layer thickness ($h_{sl}$)}
In order to verify how well the simulated surface thickness ($h_{sl}$) matches with the measured one we are forced 
to compute the typical height of the surface layer using the same criterion as in TR2008. 
The authors defined the thickness $h_{sl}$ of the surface layer as the part containing
90\% of the first kilometer boundary layer optical turbulence:
\begin{equation}
 \label{eq:bl1}
 \frac{ \int_{8m}^{h_{sl}} C_N^2(h)dh }{ \int_{8m}^{1km} C_N^2(h)dh } < 0.90
\end{equation}
where $C_N^2$ is the refractive index structure parameter.
\par
The observed $h_{sl}$ for 15 winter nights from TR2008 are
reported on table \ref{tab:hl0}. The simulated $h_{sl}$ for the same 15 nights, calculated for the 
two configurations (monomodel i.e. low horizontal resolution and grid-nesting i.e. high horizontal resolution), 
are reported in Table \ref{tab:hl1}.

\begin{table}
 \centering
  \caption{Surface layer thickness $h_{sl}$ for 15 winter nights (TR2008). Units in meter (m). 
           The criterion used is the one from Eq.\ref{eq:bl1}. The mean value is 
           also reported with the associated statistical error $\sigma$/$\sqrt{N}$ where N is the number of independent estimates.}
  \begin{tabular}{|c|c|c|c|} 
   \hline
   Date &  Observed surface  & Date & Observed surface    \\
        & layer thickness &      & layer thickness \\
   \hline
   04/07/05 & 30 & 12/08/05 & 22 \\
   07/07/05 & 21 & 29/08/05 & 47 \\
   11/07/05 & 98 & 02/09/05 & 41 \\
   18/07/05 & 26 & 05/09/05 & 20 \\
   21/07/05 & 47 & 07/09/05 & 39 \\
   25/07/05 & 22 & 16/09/05 & 24 \\ 
   01/08/05 & 40 & 21/09/05 & 22 \\
   08/08/05 & 30 & \\
   \hline
            & Mean  & 35.3 & \\
   \hline
             & $\sigma$  & 19.9 & \\
   \hline          
          & $\sigma$/$\sqrt{N}$  & 5.1 & \\
  \hline
  \end{tabular}
 \label{tab:hl0}
\end{table}
\begin{table}
 \caption{Mean surface layer thickness $h_{sl}$ for the same 15 winter nights than in table 
          \ref{tab:hl0}, deduced from Meso-Nh computations using the criterion in 
          Eq. \ref{eq:bl1}. 
          Two different times intervals were chosen: between 11 UTC and 18 UTC, 
          and between 12 UTC and 16 UTC.
          Units in meter (m). The mean value is also reported  with the associated statistical error $\sigma$/$\sqrt{N}$.}
          \begin{tabular}{|c|c|c|c|c|}
 \hline
 \multicolumn{1}{|c|}{Date} &
 \multicolumn{2}{c|}{Surface layer thickness} &
 \multicolumn{2}{c|}{Surface layer thickness} \\
 \multicolumn{1}{|c|}{} &
 \multicolumn{2}{c|}{Meso-Nh grid-nesting} &
 \multicolumn{2}{c|}{Meso-Nh monomodel} \\
 \hline
 & {\scriptsize 11-17 UTC} & {\scriptsize 12-16 UTC}
 & {\scriptsize 11-17 UTC} & {\scriptsize 12-16 UTC}
 \\ \cline{2-5}
 04/07/05 & 24.5 & 25.3 &  43.9 & 44.9  \\
 07/07/05 & 37   & 36.9 &  33   & 32.9  \\
 11/07/05 & 65.1 & 63.9 &  91.7 & 91.7  \\
 18/07/05 & 38.9 & 38.2 &  72.4 & 72.3  \\
 21/07/05 & 46.4 & 46.2 &  56.9 & 56.8  \\
 25/07/05 & 83.4 & 82.1 &  31.6 & 32    \\
 01/08/05 & 27.9 & 27.9 &  41.1 & 41    \\
 08/08/05 & 33.3 & 29.7 & 145.7 &153.3  \\
 12/08/05 & 23.1 & 18.2 &  44.3 & 44.8  \\
 29/08/05 & 92   & 84.6 &  77.3 & 78    \\
 02/09/05 & 52.8 & 54.5 &  66.5 & 67.1  \\
 05/09/05 &120   &122   &  37.2 & 38.8  \\
 07/09/05 & 58.1 & 56.9 &  89   & 85.9  \\
 16/09/05 & 27.6 & 29.4 &  40   & 41.7  \\
 21/09/05 & 17   & 17.3 & 105   &107.5  \\
 \hline
 Mean     & 49.8 & 48.9 &  65   & 65.9 \\
 \hline
 $\sigma$    & 29.4 & 29.3 & 32.7  & 33.6 \\
 \hline
 $\sigma$/$\sqrt{N}$    &  7.6 &  7.6 &  8.2  &  8.7 \\
 \hline
\end{tabular}
 \label{tab:hl1}
\end{table}
\par

\par
Two different time intervals were chosen for the computations of the mean values (11 UTC to 17 UTC 
and 12 UTC to 16 UTC), both temporally centered around 14 UTC i.e the time at which the balloons have been launched.
We reported on Figure \ref{fig:corr_slt} a correlation plot where the measured and the simulated (mean values
 between 12 UTC and 16 UTC only) $h_{sl}$ are compared.
\par
It is worth to highlight how the comparison between measurements and simulations is done. 
At present time, it is meaningless to predict the $\CN2$ at a time t$=$t$_{0}$. 
The optical turbulence is indeed a parameter fluctuating in space and time at much smaller scales than what a classical meteorological parameter does and at much smaller 
scales than the model mesh sizes. 
The optical turbulence is parameterized in the model and not resolved explicitly and it has to be quantified statistically. For this reason it is not realistic at present time, to forecast the $\CN2$ profiles with a better precision in time than a $\Delta$t of a few hours\footnote{We refer the reader to \cite{m4}-Session 2 for an extended explanation of this concept. The perspective of this typology of studies is to attain smaller and smaller $\Delta$t.}. 
Our objective is to find therefore a correlation between the measurements obtained during one balloon launch ($\CN2$ profile)
and the mean of the temporal evolution of the simulations ($\CN2$ profiles) extended on a $\Delta$t of a few hours. 
\begin{figure*}
\begin{center}
\includegraphics[width=0.9\textwidth]{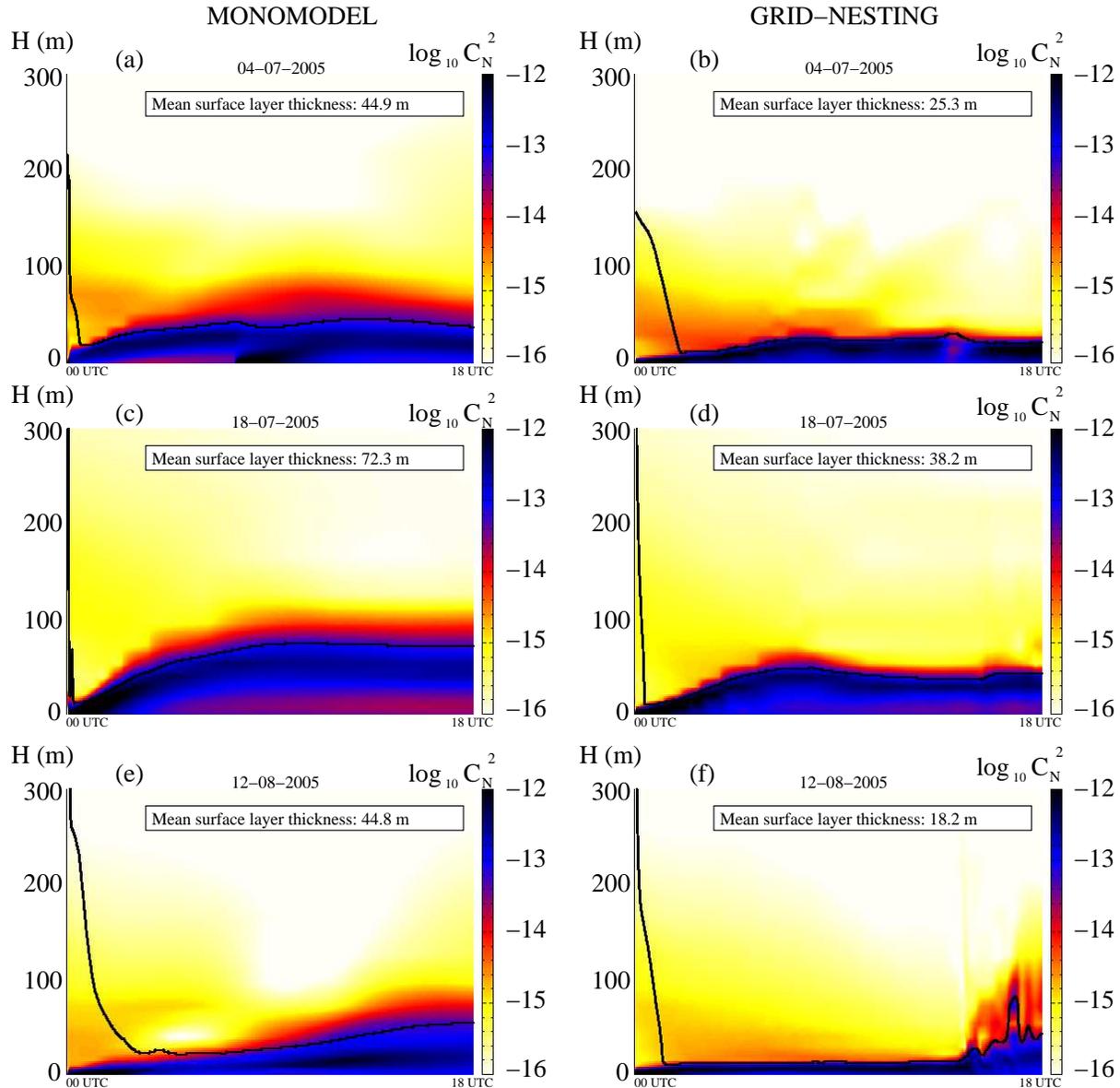}
\end{center}
\caption{Temporal evolution of the $C_N^2$ profile at Concordia Station, Dome C
         obtained with Meso-Nh, between 00 UTC and 18 UTC for the three nights:
         4/7/2005, 18/7/2005 and 12/8/2005 (from the top to the bottom).
         On the left-side are shown the simulations obtained with the low horizontal
         resolution, on the right-side the simulations obtained with the high horizontal resolution.
         The mean value of the surface layer computed with the criterion \ref{eq:bl1}
         between 12 UTC and 16 UTC is reported for each night.
         The thin black line represents the evolution of the surface layer height
         during the night. The first couple of hours can be considered as spurious values
         because of the model adaptation to the ground.}
{\label{fig:cn2}}
\end{figure*}
Looking at Table~\ref{tab:hl1} we observe that the values of the simulated surface layer thickness are weakly 
dependent on the selection of the temporal interval of integration ($\Delta$t: 11-17 UTC and 12-16 UTC). We chose the second one because
the associated simulations give a slightly better correlation with the measurements (Table~\ref{tab:hl0}). 
\begin{figure*}
\begin{center}
\includegraphics[width=0.9\textwidth]{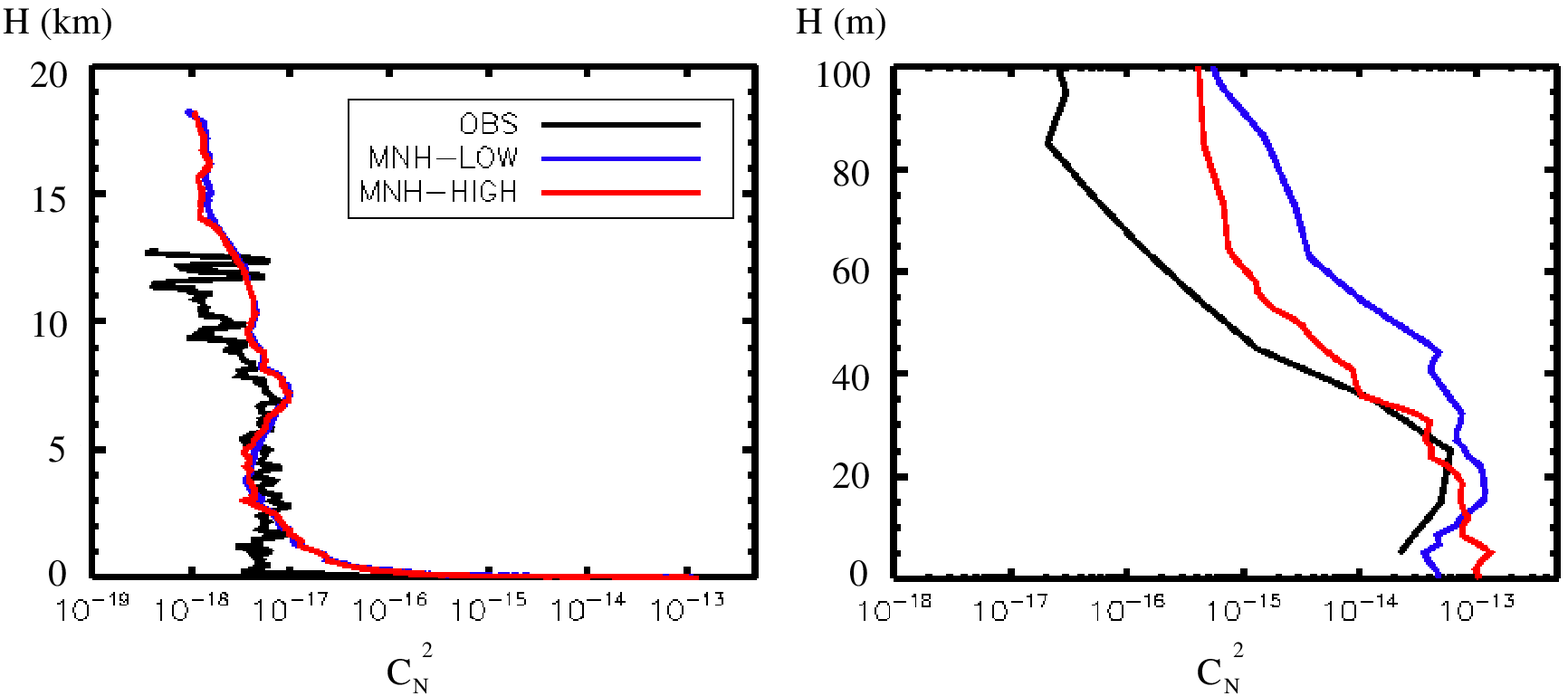}
\end{center}
\caption{{\bf Left:} Median $\CN2$ profile measured (black line) with microthermal sensors mounted on balloons (from
Trinquet et al. 2008) and simulated with the Meso-Nh mesoscale model with the low-horizontal resolution (blue line) 
and the high-horizontal resolution (red line) {\bf Right:} zoom in the first 100 m. Units in m.$^{-2/3}$.}
{\label{cn2_avg}}
\end{figure*}
Using the criterion expressed by Eq. \ref{eq:bl1},  the grid-nested simulations give a mean 
surface thickness $h_{sl,mnh-high}$$=$ 48.9$\pm$7.6~m (where $\sigma$/$\sqrt{N}$$=$7.6~m) and the monomodel a mean thickness $h_{sl,mnh-low}$$=$ 65.9$\pm$8.7~m (Table \ref{tab:hl1}).
The low resolution configuration leads to a higher mean forecasted thickness with respect to 
the observed one ($h_{sl,obs}$$=$35.3$\pm$5.1~m) while the grid-nested technique is closer to the 
observations ($\Delta$$h_{obs,mnh-high}$ $=$13.6~m) and within the typical $\sigma$. If we take into account the statistical error ($\sigma$/$\sqrt{N}$$=$7.6~m) we conclude that 
the mesoscale model provides, for this statistical sample of 15 nights, a typical surface layer thickness just $\sim$ 6 m higher than the observed one.  The dispersion $\sigma$ of the surface layer thickness for the simulated nights ($\sigma$=29.3 m) is just slightly larger than the observed one ($\sigma$=19.9 m). This indicates that the intrinsic dispersion of the $h_{sl}$ is well reconstructed by the model. Fig. \ref{fig:corr_slt} shows the correlation between simulated and observed $h_{sl}$ with the corresponding $\sigma$ values. 
We observe moreover, that, in spite of the more expensive simulations in terms of computing resources 
(time and memory), the high horizontal resolution grid-nested configuration seems 
to be necessary to better reconstruct the concentration of the turbulence in a 
thin layer near the surface. 
More precisely a horizontal resolution of 100~km  provides a bias in the typical 
$h_{sl}$ of the order of $\sim$ 30~m. 
This result is relevant with respect to the study done by SG2006 who used a resolution of 100~km and found a mean 39~m 
on a sample of 90 nights. 
Unfortunately, the authors define the $h_{sl}$ as the elevation (starting from the lowest model layer) at which the 
turbulent kinetic energy contains 1$\%$ of the turbulent kinetic energy of the lowest model layer. 
This definition is completely different from that used by TR2008 (and by us) i.e. the elevation in which is included  90\% of the optical turbulence developed in the first kilometer. It is therefore a no sense whatever comparison of measured $h_{sl}$ with the $h_{sl}$ by SG2006. 
To have an idea of the impact of the selected criteria on the mean $h_{sl}$ estimate we repeated the statistical calculation using the same criterion employed by SG2006. 
The mean $h_{sl}$ obtained with the high resolution is 32.3~m that is $\sim$ 17~m lower than the 48.9 m estimated with the criterion used by TR2008. We therefore deduce that the typical thickness estimated by SG2006 using the same criterion employed by TR2008 should be certainly much larger than 39 m. 
This confirm the fact that mesoscale models provides an overestimation of the $h_{sl}$ if they are run with a $\Delta$X$=$ 100 km. \newline
What about the morphology of the surface layer ? Looking at Table \ref{tab:hl0} 
it is well visible that the dispersion ($\sigma$) of the measurements is relatively large 
and sometimes the surface layer thickness can be of the order of many tens of meters. 
Is the mesoscale model Meso-Nh able to reconstruct such a large variability in the morphology of the surface layer ?
Table \ref{tab:hl1} shows that basically for all the nights, the model reconstructs within the $\sigma$ value the 
corresponding observed value. 
In just one case (5/9/2005) the simulated thickness is much larger ($\sim$ 120 m instead of the observed 20 m). 
We highlight that a larger surface thickness does not mean necessarily a thicker and developed turbulence near the ground but simply that 
in the (0, h$_{sl}$) range there is 90$\%$ of the turbulence developed in the first kilometer. 

To highlight the potentiality of Meso-Nh in discriminating between different 
turbulent nights, and to better visualize the impact of the model configuration (low and high resolution) on the 
forecasted optical turbulence, the temporal evolution of the $C_N^2$ 
profiles near the surface and the corresponding surface layer heights are displayed for three 
nights (Fig.~\ref{fig:cn2}). 
The first hours of the simulation should not be taken into account
because they correspond to the spin-up of the model, and do not describe realistic 
$C_N^2$ profiles. Fig.~\ref{fig:cn2}, left-side shows the simulations with low resolution, 
Fig.~\ref{fig:cn2}, right-side shows the simulations with high resolution.
The three selected nights are, from the top to the bottom: 4/7/2005, 18/7/2005 and 12/8/2005.
In all the three cases the developed turbulence layer is thinner for the high horizontal resolution case than for the low resolution one. 
The high resolution provides also a better correlation to observations. 
\begin{itemize}
\item During the night 4/7/2005  (Fig.~\ref{fig:cn2}a,b for monomodel and grid-nested 
configurations, respectively) the predicted surface layer remains constant in 
time, with a mean forecasted thickness well correlated to the observed one 
($h_{sl,mnh-high}\simeq$25.3~m, $h_{sl,mnh-low}\simeq$44.9 m and $h_{sl,obs}$$=$ 30~m). The high resolution simulation 
matches the observations within $\sigma$/$\sqrt{N}$.
\item During the night 18/7/2005  (Fig.~\ref{fig:cn2}c,d for monomodel and grid-nested 
configurations, respectively) the predicted surface layer remains constant in 
time. The simulation at low horizontal resolution ($h_{sl,mnh-low}\simeq$72.3 m) overestimates the observation ($h_{sl,obs}$$=$ 26~m) while the simulation with high resolution ($h_{sl,mnh-high}\simeq$~38.2m) is better correlated.
\item During the night 12/8/2005 (Fig.~\ref{fig:cn2}e,f for monomodel and grid-nested 
configurations, respectively) it is visible that the morphology of the reconstructed surface layer is thin as well as in the previous cases. 
For this night, the high horizontal resolution permits to put in evidence a better sensitivity of the model 
to the temporal variability of the $\CN2$ that has been observed also in other nights. Indeed we observe, just above the surface layer in the last part of the simulation, short-bumps of optical turbulence i.e. fluctuations of the $C_N^2$ profile forecasted 
by the model near the surface. This $\CN2$ variability is the signature of an evident temporal evolution of the 
turbulent energy distribution even in conditions of a strongly stratified atmosphere. 
Also in this last case, even if both simulated heights ($h_{sl,mnh-high}\simeq$18.2~m and $h_{sl,mnh-low}\simeq$44.8~m) 
are well below 100 m, the high resolution provides a clearly better correlation (within $\sigma$/$\sqrt{N}$) to the observations ($h_{sl,obs}$$=$ 22~m).
\end{itemize}
\subsection{Optical turbulence vertical distribution: seeing in the free atmosphere and in the whole atmosphere}
\par
Figure~\ref{cn2_avg} shows the median of the $\CN2$ profile measured by the microthermal sensors mounted on the balloons (15) launched at Dome C during the winter 2005 and simulated by the Meso-Nh model with the low and high horizontal resolutions. 
We observe that the shape of the $\CN2$ is well reconstructed all along the 13~km by the model. 
Also the model can reconstruct the $\CN2$ above 13~km. 
In this region a comparison with measurements is not possible because the balloons usually 
explode at these heights (see discussion in \cite{ha}). 
In the first kilometer, the simulated $\CN2$ profile decreases (from the ground to higher altitudes) in a less 
sharp way than what monitored by observations (at $\sim$ 30~m from the ground). 
This is not surprising and it derives from the fact that simulations are slightly less thermally stable than the 
observations near the surface. 
However, in the zoom of the first 100~m (Fig.~\ref{cn2_avg}, right-side) it is well visible that the shape of the 
median simulated $\CN2$ profile done with the high horizontal resolution configuration is much better 
correlated with the shape of the observed median $\CN2$ profile than the one obtained with 
a low horizontal resolution. \newline
The seeing in the free atmosphere and in the whole atmosphere for $\lambda$$=$0.5$\times$10$^{-6}$m are:
\begin{equation}
\varepsilon_{FA}=5.41 \cdot \lambda^{-1/5} \cdot \left( \int_{h_{sl}}^{h_{top}} C_N^2(h) \cdot dh \right) ^{3/5}
\end{equation}
\begin{equation}
\varepsilon_{TOT}=5.41 \cdot  \lambda^{-1/5} \cdot \left( \int_{8m}^{h_{top}} C_N^2(h) \cdot dh \right) ^{3/5}
\end{equation}
with h$_{top}$ $\sim$ 13 km from the sea level i.e. where the balloons explode and we have no more their signal.
\par
\begin{figure}
\begin{center}
\includegraphics[width=0.35\textwidth]{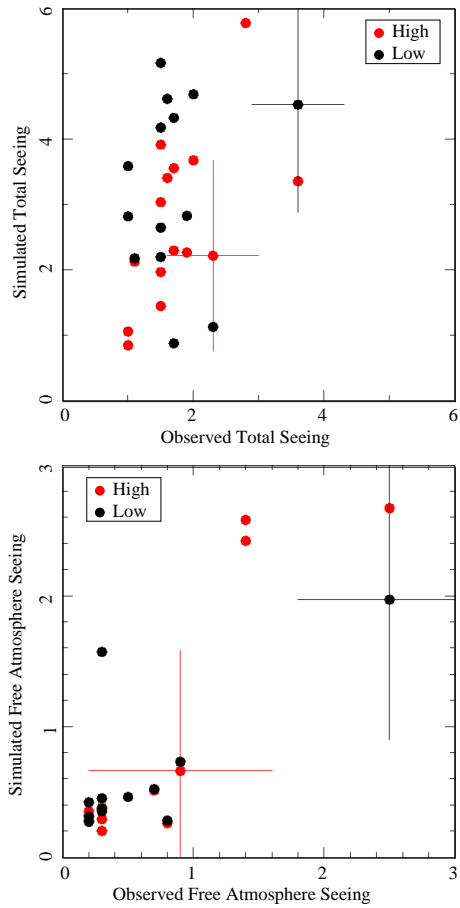}
\end{center}
\caption{Correlation plot between measured and simulated total (top) and free atmosphere (bottom) seeing 
(black: monomodel configuration; red: grid-nested configuration). 
For the simulated values only the mean values between 12 UTC and 16 UTC are considered. 
For each configuration of the simulation (high and low horizontal resolution) the error bars are 
reported for one point only (and are equal to $\sigma$). Units are in arcsec.}
{\label{fig:corr_see}}
\end{figure}

Table~\ref{see_fa} reports, for the 15 nights, the seeing in the free atmosphere ($\varepsilon_{FA}$$=$$\varepsilon_{[h_{sl},h_{top}]}$) and in the whole atmosphere ($\varepsilon_{TOT}$$=$$\varepsilon_{[8m,h_{top}]}$) calculated for the simulations and the observations. Values of the observed $\varepsilon_{TOT}$ are taken from TR2008. We considered h$_{sl}$$=$ 33m instead of 35.3 m (Table~\ref{tab:hl0}) for the observations because, in the Trinquet et al. (2008) paper, the authors provide the seeing integrated above h$_{sl}$, where h$_{sl}$ is  calculated on all the nights belonging to the [February-November] range. We considered for the simulations h$_{sl}$ as retrieved from Table~\ref{tab:hl1}.

\begin{table}
\caption{Total seeing $\varepsilon_{TOP}$$=$$\varepsilon_{[8m,h_{top}]}$ and seeing in the free atmosphere $\varepsilon_{FA}$$=$$\varepsilon_{[h_{sl},h_{top}]}$ calculated for the 15 nights and averaged in the temporal range 12-16 UTC. See the text for the definition of h$_{sl}$ and h$_{top}$. In the second column are reported the observed values, in the third and fourth columns the simulated values obtained with high and low horizontal resolution respectively. Units in arcsec.}
\begin{tabular}{cccc}
 \hline
           & Obs.        & MESO-NH              & MESO-NH                \\
           &             & HIGH                 & LOW                \\
 \hline
  Date     & $\varepsilon_{FA}$/$\varepsilon_{TOT}$      & $\varepsilon_{FA}$/$\varepsilon_{TOT}$      & $\varepsilon_{FA}$/
$\varepsilon_{TOT}$        \\
           & {\tiny($h_{sl}$=33m)} & {\tiny($h_{sl}$=48.9m)} & {\tiny($h_{sl}$=65.9m)} \\
 \hline
  04/07/05 &      0.3 / 1.6       &     0.20 / 3.40      &    0.45 / 4.61         \\
  07/07/05 &      0.2 / 1.5       &     0.35 / 3.03      &    0.31 / 2.64         \\
  11/07/05 &      1.4 / 1.7       &     2.42 / 3.55      &    3.27 / 4.32         \\
  18/07/05 &      0.3 / 2.0       &     0.35 / 3.67      &    1.57 / 4.68         \\
  21/07/05 &      0.7 / 1.1       &     0.51 / 2.12      &    0.52 / 2.17         \\
  25/07/05 &      0.3 / 1.0       &     0.29 / 0.84      &    0.35 / 3.58         \\
  01/08/05 &      0.8 / 1.6       &     0.26 / 3.91      &    0.28 / 4.17         \\
  08/08/05 &      0.5 / 2.3       &     0.46 / 2.21      &    0.46 / 1.12         \\
  12/08/05 &      0.2 / 1.5       &     0.32 / 1.96      &    0.42 / 5.16         \\
  29/08/05 &      2.5 / 3.6       &     2.67 / 3.35      &    1.97 / 4.52         \\
  02/09/05 &      0.9 / 1.9       &     0.66 / 2.26      &    0.73 / 2.82         \\
  05/09/05 &      0.3 / 1.0       &     0.38 / 1.05      &    0.37 / 2.81         \\
  07/09/05 &      1.4 / 2.8       &     2.58 / 6.77      &    3.33 / 7.02         \\
  16/09/05 &      0.2 / 1.5       &     0.28 / 1.44      &    0.27 / 2.19         \\
  21/09/05 &      0.2 / 1.7       &     0.32 / 2.29      &    0.27 / 0.87         \\
 \hline
  Median   &      0.3 / 1.6       &     0.35 / 2.29      &    0.42 / 3.58         \\
 \hline
  $\sigma$ &      0.7 / 0.7       &     0.92 / 1.46      &    1.07 / 1.64         \\
 \hline
  $\sigma$/$\sqrt{N}$ & 0.2 / 0.2 &     0.24 / 0.38      &    0.28 / 0.42         \\
 \hline
\end{tabular}
\label{see_fa}
\end{table}
Again we observed that results are weakly dependent on the temporal range on which the means values are calculated 
and for this reason we report just the 12-16 UTC case. Fig. \ref{fig:corr_see} shows the correlation between the observed and simulated 
values for the seeing in the free atmosphere and in the whole atmosphere.
The median of the observed seeing in the free atmosphere for the 15 nights is $\varepsilon_{FA,obs}$$=$ 0.3$\pm$0.2 
arcsec; the median seeing in the free atmosphere simulated by Meso-Nh with the high horizontal resolution 
is $\varepsilon_{FA,mnh-high}$$=$ 0.35$\pm$0.24 arcsec and with the low horizontal resolution 
is $\varepsilon_{FA,mnh-high}$$=$ 0.42$\pm$0.28 arcsec. 
Both median simulated values (with low and high resolution) match the median value obtained with observation within 
the statistical error even if the high resolution is much better correlated 
(relative error of 16$\%$, $\Delta$$\varepsilon_{obs,sim}$$=$0.05"). 
If we look at the total seeing developed on the whole atmosphere it is well visible (Table~\ref{see_fa} and Fig.\ref{fig:corr_see}) that the 
model overestimates the measurements with both resolutions. We have a simulated median $\varepsilon_{TOT,mnh-low}$$=$ 3.58$\pm$0.42 arcsec and $\varepsilon_{TOT,mnh-high}$$=$ 2.29$\pm$0.38 arcsec versus an observed median $\varepsilon_{TOT,obs}$$=$ 1.6$\pm$0.2 arcsec. Even if we take into account the more accurate estimates (high resolution) we obtain a dispersion simulations/observations $\Delta$$\varepsilon$ $\sim$ 0.7 arcsec. The excess of optical turbulence reconstructed by the Meso-Nh model is clearly concentrated in the surface layer. We can not exclude an underestimate from measurements but there are, at present time, no major elements that lead to this assumption. We are working, on the contrary, on a paper to explain this discrepancy and overcome this limitation. Considering that we proved that the meteorological parameters are well reconstructed by the Meso-Nh model near the surface (Section 3) and that the surface numerical scheme (Interaction
 Soil Biosphere Atmosphere ISBA) responsible of the control of the budget of the turbulent ground/air fluxes has been 
recently optimized for Antarctic applications (\cite{lemoigne08}) in the context of our project, we concentrated our attention on the dynamical and optical numerical turbulence schemes. 

In terms of comparison with the SG2006 study we note that  the latter study indicates a typical underestimated total 
seeing of 1.16 arcsec with respect to the observed one (1.6 arcsec). 
The discrepancy is smaller from a quantitative point of view ($\Delta$$\varepsilon_{TOT}$$=$0.45 arcsec) with respect to 
what we find and it is in the opposite direction. 
The questionable issue in the SG2006 study is that the turbulence kinetic energy provided by SG2006 in the first levels 
of the MAR model is often of the order of 10$^{-4}$ m$^{2}$s$^{-2}$ (\cite{gallee2007}). 
This values is extremely low and it basically indicates no turbulent kinetic energy on the first level of the model and such a condition is contrary to what observed with measurements. This is consistent with the fact that the MAR model underestimates the seeing in the surface layer. 
\par
\begin{figure*}                                                                                                           
\begin{center}                                                                                                            
\includegraphics[width=0.9\textwidth]{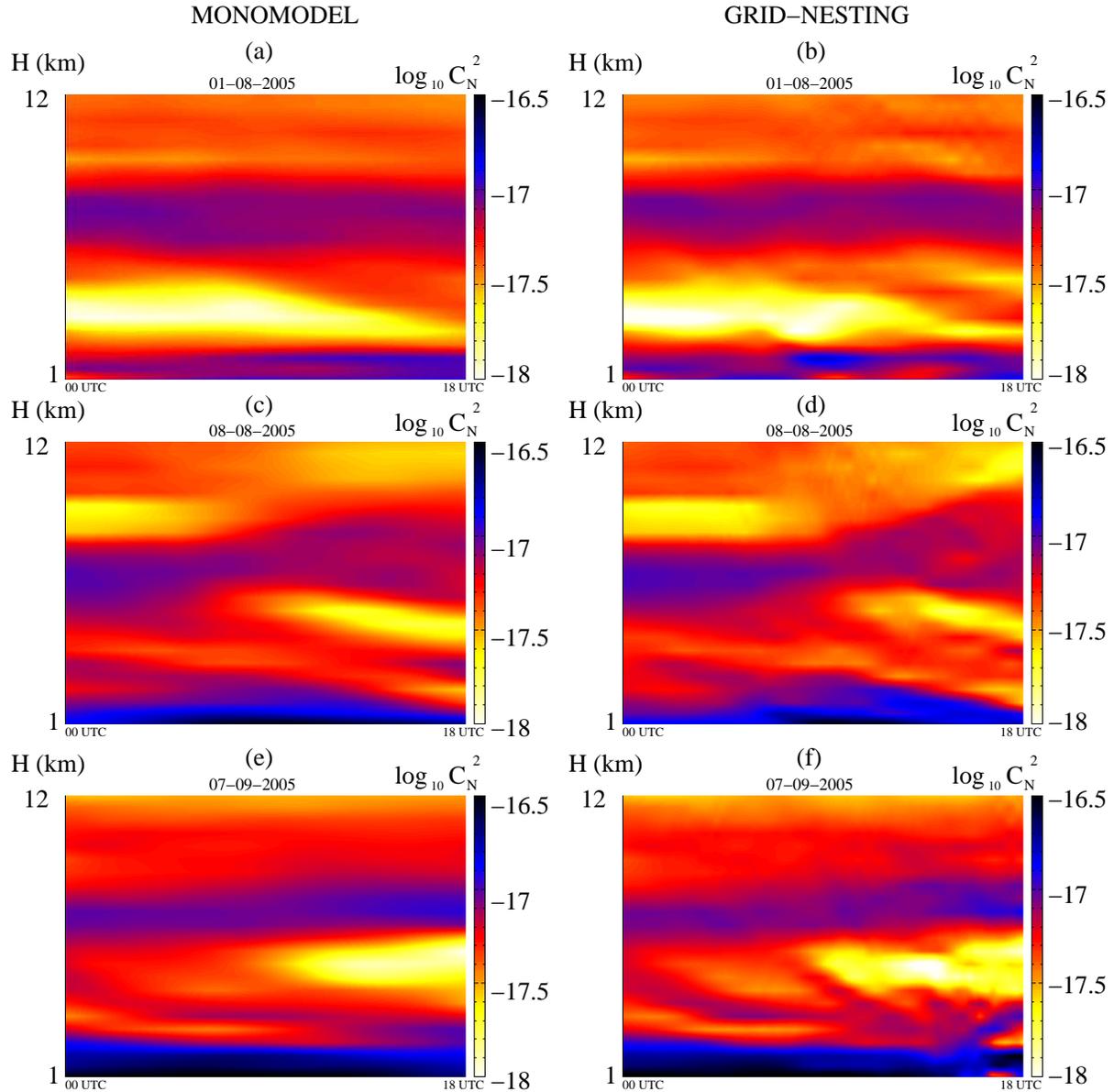}
\end{center}                                                                                                              
\caption{Temporal evolution on 18 hours of the $\CN2$ profiles in the vertical slab (1-12) km related to three nights
 chosen in the selected sample of 15 nights. The dynamic of the $\CN2$ (color palette) has been tuned to put in evidence the variation of the $\CN2$ values (in logarithmic scale) during the time in the range (-18,-16.5).}                            
{\label{see_free}}                                                                                                        
\end{figure*}

We conclude that the Meso-Nh model, in the present configuration, reconstructs with good statistical reliability the $h_{sl}$ and the seeing in the free atmosphere while shows a tendency in overestimating the strength of the seeing in the surface layer. The interesting result  of this paper is therefore the fact that the most important features for astronomical interest (the surface layer thickness and the typical seeing in the free atmosphere) observed with measurements are confirmed with mesoscale atmospherical model.  We note that the this is the first confirmation made by a mesoscale model of the typical seeing in the free atmosphere.
Besides it is worth to highlight that these are the first $\CN2$ simulations ever done above the internal Antarctic Plateau 
and extended all along the whole atmosphere. 
Figure~\ref{see_free} shows the temporal evolution of the $\CN2$ profile in the free atmosphere 
(more precisely in the (1,12) km vertical slab) 
related to three selected nights in the sample of the 15 simulated nights. 
In all of the three nights it is well visible that, even at such high altitudes, 
the model is active and the vertical distribution of the optical turbulence changes in time with a not 
negligible dynamic from a quantitative point of view. 
The $\CN2$ values extend, indeed, on the logarithmic scale (-18,-16.5). 
In all the 3 cases it appears clearly that the high-horizontal resolution provides a better temporal 
variability as expected. 
These results are therefore very promising in terms of predictions of the $\CN2$ 3D maps on long time scales. 
\section{Conclusions}

In this paper we study the performances of the Meso-Nh mesoscale meteorological 
model in reconstructing meteorological parameters (wind speed and temperatures) as well 
as the optical turbulence above Concordia Station in the Dome C area, a site in the internal Antarctic Plateau. 
This is, at our knowledge, the first study concerning the optical turbulence reconstructed with an atmospherical mesoscale model above Antarctica on the whole atmosphere.
This study is concentrated on the winter season i.e. the most interesting for stellar astronomical applications. 
The validation of the model for the meteorological parameters has been done comparing measurements (radiosoundings) and simulations on a sample of 47 nights. 
The validation of the model for the optical turbulence has been done comparing simulations with measurements on a sample of 15 nights.
Two different model configurations were tested: monomodel simulations using a low horizontal resolution (${\Delta}X$=100~km)
and grid-nesting simulations with high horizontal resolution (${\Delta}X$=1~km for the innermost domain). The low resolution model permitted us to discuss the results obtained previously in the  literature.
The observations used for the validation are, for the meteorological parameters, the analyses from the ECMWF Global Circulation Model
and radiosoundings (47 nights) and, for the optical turbulence, the $\CN2$ and seeing values (15 nights) measured in situ (Trinquet et al. 2008). 
\newline

The main conclusions of this study are:
\begin{itemize}
\item[(1)] We showed that near the surface, Meso-Nh retrieved better wind speed
vertical gradient (wind shear) than the ECMWF analyses from a qualitative as well as quantitative point of view, thanks to the use of a highest vertical 
resolution. We expect therefore a better reconstruction of the katabatic winds typical of these regions by the Meso-Nh model than the GCM models. 
Also Meso-Nh better reconstructs the thermal stability near the surface than the GCMs.
The analysis of the first vertical grid point permits us to conclude 
that the Meso-Nh model surface temperature is closest to the observations $\Delta$T$_{mnh-high,obs}$$=$1.60 K 
than the ECMWF General Circulation Model ($\Delta$T$_{ecmwf,obs}$$=$3.74 K) which is too warm. 
The improvement for the estimate of the wind speed is even more evident ($\Delta$V$_{mnh-high,obs}$$=$
0.04~m.s$^{-1}$ versus $\Delta$V$_{ecmwf,obs}$$=$ 2.49~m.s$^{-1}$). \\
\item[(2)] For what concerns the parameters concerning the optical turbulence, again the results are resolution dependent. The simulations with low resolution provides a too thick surface layer (almost double of the observed one) while those with high resolution provide a mean h$_{sl,mnh-high}$$=$48.9$\pm$7.6~m versus an equivalent observed h$_{sl,obs}$$=$35.3$\pm$5.1~m.  
Taking into account the statistical error we observe that the high horizontal mode provides a surface layer thickness that is statistically just 6 m higher than the observed one but within the dispersion $\sigma$ of the observations.\\
\item[(3)] The integral of the $\CN2$ above the h$_{sl}$ i.e. the seeing in the free atmosphere $\varepsilon_{FA,obs}$$=$ 0.3$\pm$0.20 arcsec is reconstructed with an excellent level of reliability ($\Delta$$\varepsilon$$=$0.05 arcsec) by the model used with the high resolution configuration $\varepsilon_{FA,mnh-high}$$=$ 0.35$\pm$0.24 arcsec. The low resolution provides a worse estimate even if within the $\sigma$ of the observations. \\
\item[(4)] The model still shows a tendency in overestimating the turbulence in the surface layer. For an observed 
$\varepsilon_{TOT,obs}$$=$ 1.6 arcsec we have a simulated $\varepsilon_{TOT,mnh}$$=$ 2.29 arcsec with the model in high horizontal resolution mode. This is the subject of an on-going study conceived to answer to this open question.\\
\item[(5)] The results concerning the computation of the mean thickness 
of the surface layer as well as the seeing in different vertical slabs are not very dependent of the time interval used to average 
it. This widely simplifies the analysis of simulations.\\
\item[(6)] Estimates obtained with the grid-nested simulations are closer to the 
observations than those obtained with monomodel simulations.
This study highlighted the necessity of the use of high horizontal resolution to 
reconstruct a good meteorological field as well as the parameters characterizing the optical turbulence in Antarctica, 
even if the orography is almost flat over the internal Antarctic Plateau. 
The employment of the low resolution (100 km) alone can hardly be used to identify {\it the best site on the Antarctic Plateau}.  However, it can be used to identify rapidly, on the whole Antarctic Plateau, the most interesting regions in which to focus, successively, simulations at high horizontal resolutions on smaller surfaces domains. With "the most interesting regions" we mean those with the lowest surface layer thickness for example.  \\
\item[(7)] The Meso-Nh model is able to reconstruct a mean $\CN2$ profile well fitting the vertical optical turbulence distribution measured in the first 20 km from the ground. The model also shows a not negligible temporal variability in the whole 20 km from the ground in a very small dynamic range. The latter is to be considered a very interesting feature because it is known that this is a region of the atmosphere in which in general the mesoscale models are less sensible than near the ground. It is therefore a further indication that the Meso-Nh model is well placed to forecast the turbulence evolution at these time scales.\\
\end{itemize}
Once the tendency in overestimating the strength of the turbulence in the surface layer will be solved (forthcoming paper) we plan to run the Meso-Nh model in other regions of the internal 
Antarctic Plateau to identify the best locations for astronomical observations i.e. the places with the best 
turbulence characteristics from an astronomical point of view.

\section*{Acknowledgments}
This study has been funded by the Marie Curie Excellence Grant (FOROT) - MEXT-CT-2005-023878.
ECMWF analyses are extracted from the Catalog MARS, 
$http://www.ecmwf.int$.
Radiosoundings come from the Progetto di Ricerca "Osservatorio Meteo Climatologico"
of the Programma Nazionale di Ricerche in Antartide (PNRA), 
{\it http://www.climantartide.it}.


\label{lastpage}
\end{document}